\documentclass{article}
\usepackage{ijcai26}

\usepackage{times}
\usepackage{soul}
\usepackage{url}
\usepackage[hidelinks]{hyperref}
\usepackage[utf8]{inputenc}
\usepackage[small]{caption}
\usepackage{graphicx}
\usepackage{amsmath}
\usepackage{amsthm}
\usepackage{booktabs}
\usepackage{algorithm}
\usepackage{algorithmic}
\usepackage[switch]{lineno}

\usepackage[center]{subfigure}
\usepackage{multirow}
\usepackage{amssymb} 

\newtheorem{definition}{Definition}

\title{FlexStructRAG: Flexible Structure-Aware Multi-Granular Relational Retrieval for RAG}

\author{
Mengzhu Chen$^1$
\and
Haodong Yang$^1$\and
Jia Cai$^{1,2}$\And
Xiaolin Huang$^{3,4}$\\
\affiliations
$^1$School of Statistics and Data Science, Guangdong University  of Finance $\&$ Economics, Guangzhou, China\\
$^2$Guangdong Provincial Key Laboratory of Public Finance and Taxation with Big Data Application,  Guangdong University  of Finance $\&$ Economics, Guangzhou, China\\
$^3$Institute of Image Processing and Pattern Recognition, Shanghai Jiao Tong University, Shanghai, China\\
$^4$MOE Key Laboratory of System Control and Information Processing, Shanghai Jiaotong University, Shanghai, China\\
\emails
hyhpchenmengzhu@gmail.com,
haodongyang@student.gdufe.edu.cn
jiacai1999@gdufe.edu.cm,
xiaolinhuang@sjtu.edu.cn
}

\begin{document}

\maketitle

\begin{abstract} Retrieval-Augmented Generation (RAG) systems critically depend on how external knowledge is segmented, structured, and retrieved. Most existing approaches either retrieve fixed-length text chunks, which fragments discourse context, or commit to a single structured index (e.g., a knowledge graph or hypergraph), which hard-codes one relational granularity. This often yields brittle retrieval when queries require different forms of evidence, such as local binary relations, higher-order interactions, or broader document-grounded context. We propose \textbf{FlexStructRAG}, a flexible structure-aware RAG framework that supports \emph{multi-granular, query-adaptive retrieval} over heterogeneous knowledge representations. FlexStructRAG jointly constructs (i) a knowledge graph for binary relations, (ii) a knowledge hypergraph for n-ary relations, and (iii) structure-aware semantic clusters that aggregate relational evidence into document-grounded context units. To reduce semantic fragmentation induced by uniform chunking, we introduce dynamic partitioning and a truncated sliding-window extraction mechanism that incorporates bounded contextual dependencies during knowledge construction. At inference time, FlexStructRAG enables entity-, edge-, hyperedge-, and cluster-level retrieval, which can be flexibly combined to supply generation with relationally and contextually aligned evidence. Experiments on the UltraDomain benchmark across four domains show that FlexStructRAG improves semantic evaluation over strong RAG baselines. Ablation and sensitivity analysis further demonstrate the necessity of multi-granular relational retrieval and structure-aware clustering. 
\end{abstract}

\section{Introduction}\label{intro}

\begin{figure}[!tb]
   \centering
	\subfigure[Graph- and hypergraph-based RAG answer only one of the two queries, while FlexStructRAG retrieves evidence for both.]{
		\centering
		\includegraphics[scale=0.26]{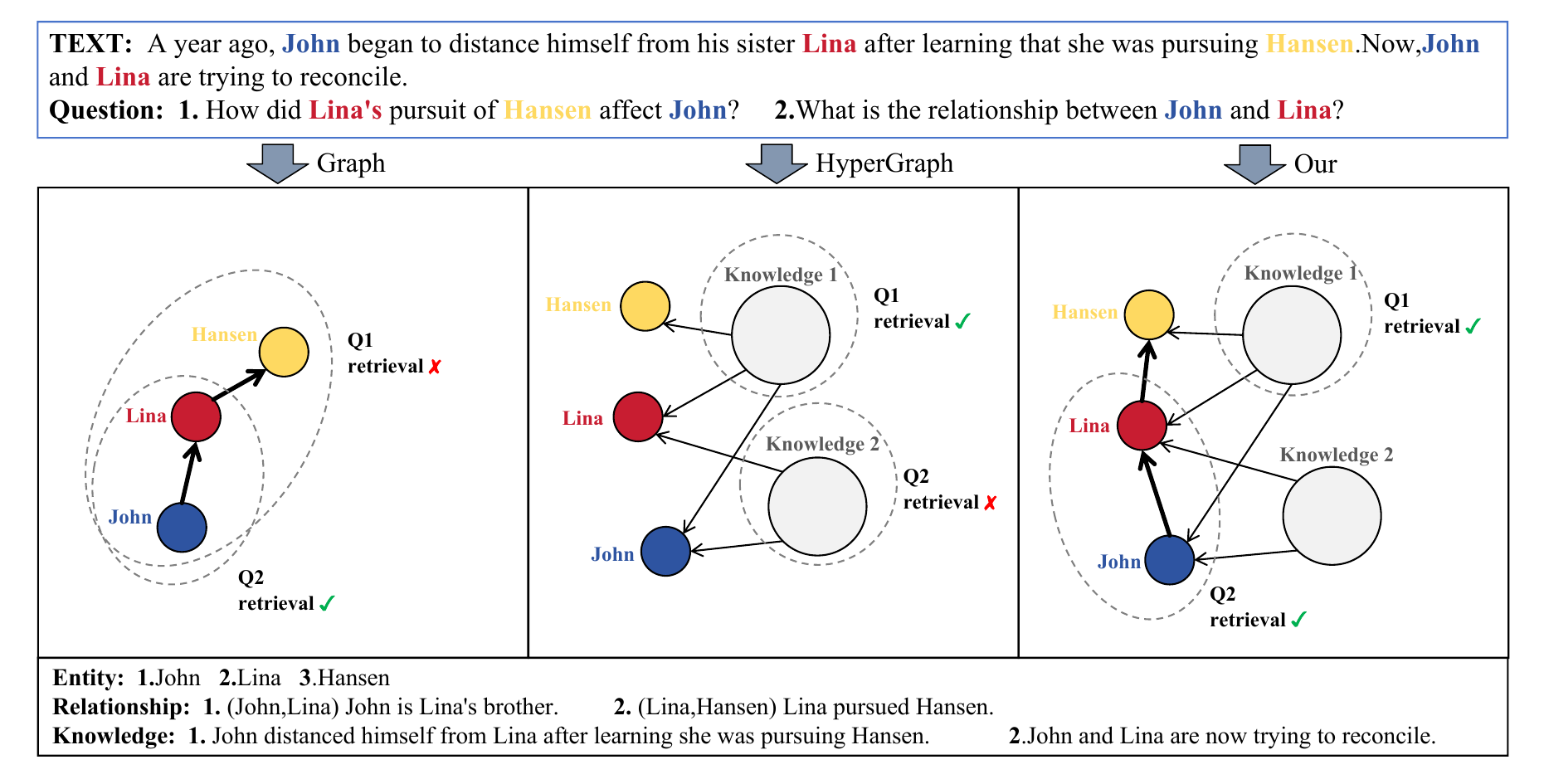}
		\label{illu1}
	}
	\subfigure[Text-only clustering may miss relational evidence or return irrelevant context; FlexStructRAG retrieves relationally aligned evidence consistently.]{
		\centering
		\includegraphics[scale=0.44]{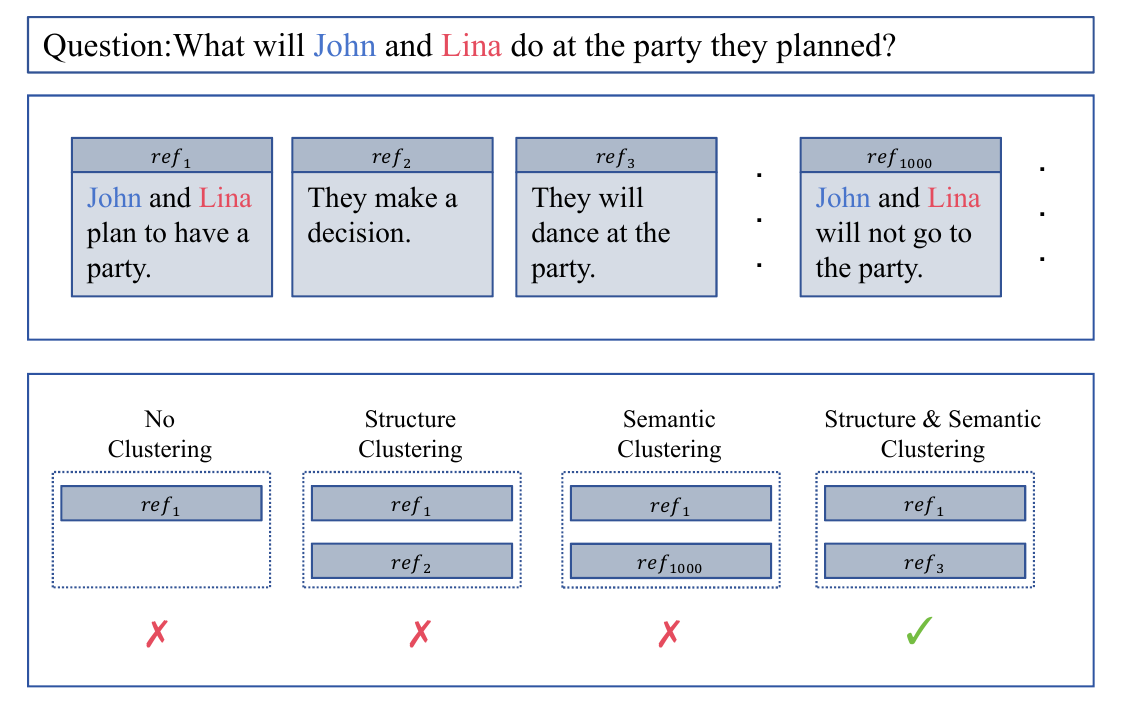}
		\label{illu2}
	}    
    \caption{Query-adaptive structural granularity in FlexStructRAG. (a) Graph-only and hypergraph-only RAG exhibit complementary failure modes due to fixed relational bias; FlexStructRAG selects the appropriate granularity to answer both query types. (b) Text-only semantic clustering can miss or dilute relational evidence; FlexStructRAG retrieves relationally aligned evidence and document-grounded context consistently.}
	\label{illu-exam}
\end{figure}

Retrieval-Augmented Generation (RAG)~\cite{lewis2020retrieval} combines external retrieval with large language models (LLMs)~\cite{achiam2023gpt,dong2022survey,gao2023retrieval,yang2025qwen3} to improve factuality and mitigate hallucinations. However, many RAG systems still retrieve fixed-length text chunks, which can fragment discourse context and weaken relational evidence required for multi-hop reasoning. Structured retrieval partially addresses this limitation by explicitly modeling entities and relations.

Graph-based RAG methods~\cite{pan2024unifying,dong2025effiqa} represent knowledge as entity--relation graphs, enabling relational selection and structured reasoning, but standard graphs primarily encode binary relations. Hypergraph-based RAG methods (e.g., HyperGraphRAG~\cite{luo2025hypergraphrag,feng2025hyper}) represent n-ary relations as hyperedges and improve performance on queries requiring higher-order structure. As illustrated in Figure~\ref{illu1}, graph- and hypergraph-based methods succeed on complementary subsets of queries: binary relations are ubiquitous and often sufficient, whereas some queries require explicit n-ary interactions. Meanwhile, hierarchical text aggregation methods such as RAPTOR cluster chunks by semantic similarity to build multi-level context, but they do not model symbolic relational structure. As a result, clustering alone can miss or dilute the relational evidence needed for precise multi-hop reasoning (Figure~\ref{illu2}). Overall, existing approaches typically hard-code a single granularity (binary, n-ary, or purely semantic), limiting robustness across diverse query intents.

\paragraph{Motivation and Novelty.}
Knowledge-intensive queries vary in the form of evidence they require. Some are resolved by localized binary relations; others require explicit n-ary interactions; and many benefit from discourse-contiguous context to bridge distant facts or resolve ambiguity. This motivates \textbf{FlexStructRAG}, which treats retrieval units at multiple granularities---\emph{entities}, \emph{edges}, \emph{hyperedges}, and \emph{document-grounded clusters}---as first-class objects within a unified framework. This design supports switching retrieval granularity at inference time without rebuilding the index. 

FlexStructRAG constructs a knowledge graph (binary relations) and a knowledge hypergraph (n-ary relations) from LLM-extracted relational items. Each structured unit retains span-level provenance linking it to supporting source text, which preserves document grounding and mitigates distortion from chunk boundaries. To ensure efficiency, we combine dynamic partitioning with truncated sliding-window extraction; this incorporates bounded cross-chunk context while maintaining linear-time complexity and cacheability.

FlexStructRAG further introduces a \textbf{meso-level structure-aware semantic clustering (SSC)} mechanism that aggregates relational primitives into intermediate, document-grounded semantic units. Here, ``meso-level'' denotes a scale between (i) micro-level primitives (a single entity/edge/hyperedge) and (ii) macro-level topic or document clusters. Concretely, SSC clusters \emph{hyperedges} that are semantically related \emph{and} close in local discourse (via positional proximity). This design restores discourse-contiguous context for multi-hop reasoning while preserving relational specificity.

In summary, FlexStructRAG unifies four retrieval unit types (entities/edges/hyperedges/clusters), performs hyperedge-centered clustering with locality, uses bounded and cacheable extraction, and supports controlled ablations via feature toggles (graph-only, hypergraph-only, and hybrid variants). Detailed comparisons to related work are provided in Appendix~B and Table~\ref{compa} in Appendix~F.

With these distinctions in place, our contributions are: 
\begin{itemize}
\item \textbf{Multi-granular retrieval units.} 
We unify entity-, edge-, hyperedge-, and cluster-level retrieval within a single framework, enabling inference-time control over retrieval granularity.
\item \textbf{Hyperedge-Centered SSC.}
We package discourse-contiguous, relationally specific evidence by clustering hyperedges using semantic similarity and positional locality.
\item \textbf{Efficient, Provenance-Preserving Construction.}
We combine dynamic partitioning and truncated sliding-window extraction with span-level provenance links for bounded, cacheable knowledge construction.
\item \textbf{Instantiation and Empirical Validation.}
We instantiate FlexStructRAG and demonstrate consistent improvements across domains via extensive experiments, ablations, and sensitivity analysis.
\end{itemize}

\section{Preliminaries}\label{prel}
We introduce graph-based RAG and hypergraph representations.
\begin{definition}[Graph-based RAG]
Graph-based RAG represents structured knowledge as a graph $G=(V,E)$, where $V$ is the set of entities (nodes) and $E$ is the set of relations (edges). A graph knowledge item can be written as $k_g=(e, V_e)$, where $e\in E$ and $V_e\subseteq V$ denotes the entities incident to $e$ (typically $|V_e|=2$).

Given a query $q$, the RAG process produces an answer $y$ by marginalizing over retrieved knowledge items:
\begin{equation*} P(y \mid q)=\sum_{k_g \in \mathcal{K}(G)} P(y \mid q, k_g)\, P(k_g \mid q, G),
\end{equation*}
where $\mathcal{K}(G)$ denotes the set of candidate knowledge items derived from $G$.
\end{definition}

\begin{definition}[Hypergraph]
A hypergraph $G^H=(V,E^H)$ generalizes a standard graph by allowing each hyperedge $e^H\in E^H$ to connect two or more entities. Each hyperedge is associated with an incident entity set
\[V_{e^H}\subseteq V,\qquad |V_{e^H}|\ge 2.\]
\end{definition}
While standard graphs are limited to binary relations, hypergraphs naturally represent n-ary relations. A hypergraph knowledge unit can be written as $k_{g^H}=(e^H,V_{e^H})\in G^H$.

\begin{figure*}[!tb]
	\centering
	\includegraphics[scale=0.33]{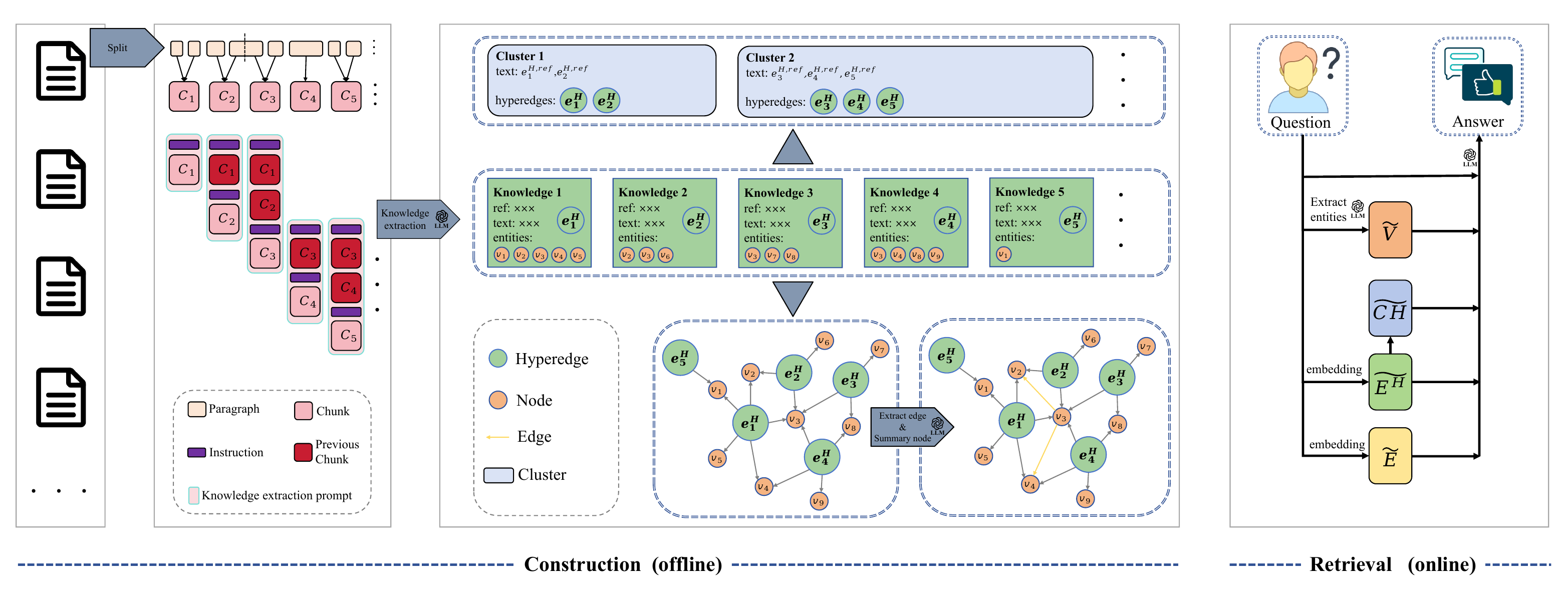}
	\caption{Overview of FlexStructRAG. Phase 1 (offline) performs dynamic document chunking and truncated sliding-window extraction, then constructs a knowledge hypergraph, a knowledge graph, and structure-aware semantic clusters. Phase 2 (online) retrieves entities, edges, hyperedges, and clusters to assemble query-aligned evidence for generation.}
	\label{framework}
\end{figure*}
\section{The FlexStructRAG Architecture}\label{method}
FlexStructRAG targets realistic, noisy knowledge extraction, where LLM outputs can be incomplete and may introduce spurious or inconsistent entities and relations.  Rather than assuming a single perfect structure, FlexStructRAG maintains complementary representations at multiple granularities---\emph{entities}, \emph{binary relations}, \emph{n-ary relations}, and \emph{document-grounded semantic clusters}---to support robust retrieval when any individual primitive is missing or corrupted. The framework comprises two stages: a knowledge-base construction phase and a retrieval phase (Figure~\ref{framework}). 

\subsection{Knowledge Base Construction Phase}
\paragraph{Document Set Processing.}
Consider two documents that both mention ``Apple''. Without sufficient context, ``Apple'' may refer to the fruit or the company, and naive linking based on lexical overlap can introduce spurious connections~\cite{panda-etal-2024-holmes}. To reduce such errors, FlexStructRAG processes each document independently and merges structured representations only after per-document construction.

Let $\mathcal{D}=\{D_1,D_2,\ldots,D_{N_D}\}$ denote the document set. For each document $D_i$, we perform chunking and knowledge extraction locally to construct a document-specific hypergraph, graph, and cluster set. When the document index is not needed for clarity, we omit the subscript $i$.

\paragraph{Dynamic Partitioning for Document Chunking.} 
Instead of using fixed-size chunks, we apply dynamic partitioning with minimum and maximum token thresholds $ct_{\min}$ and $ct_{\max}$. We first segment a document into paragraphs $\{P_j\}$ and then aggregate consecutive paragraphs into chunks $\{C_i\}$. Paragraphs are appended to the current chunk until adding the next paragraph would exceed $ct_{\max}$. If the resulting chunk is shorter than $ct_{\min}$, we merge it with the adjacent chunk in a post-processing step (without re-partitioning earlier chunks). Exceptionally long paragraphs are recursively subdivided. This procedure yields token-balanced chunks while avoiding costly global optimization.

\paragraph{Truncated Sliding-Window Knowledge Extraction.} 
Uniform chunk-level extraction can miss cross-chunk dependencies (e.g., coreference, implicit arguments, or deferred definitions). To incorporate limited context without exceeding the LLM context window, we adopt a truncated sliding-window strategy. Let $\{C_1,\ldots,C_{N_C}\}$ be the dynamically constructed chunks. For each chunk $C_i$, we define a contextual prefix $\mathrm{PreC}_i$ that contains a bounded number of preceding chunks. We first compute a window start index
 \[ 
\tau(i)=\Big\lfloor \frac{i-1-g_{\mathrm{overlap}}}{g_{\max}-g_{\mathrm{overlap}}} 
\Big\rfloor (g_{\max}-g_{\mathrm{overlap}}) + 1, 
\] 
where $g_{\max}$ is the maximum window size and $g_{\mathrm{overlap}}<g_{\max}$ controls overlap between consecutive windows. The effective start index is $s(i)=\max(1,\tau(i))$, and the prefix is \[ \mathrm{PreC}_i = \begin{cases} \emptyset, & i=1,\\ \left( C_{s(i)}, C_{s(i)+1}, \ldots, C_{i-1} \right), & i>1. \end{cases} \] 
An example with $g_{\max}=3$ and $g_{\mathrm{overlap}}=1$ is shown in Figure~\ref{framework}.  For each chunk $C_i$, we additionally retrieve up to $g_{\max}-1$ preceding chunks, with adjacent windows overlapping by $g_{\mathrm{overlap}}$ chunks to mitigate boundary information loss. This design produces token-balanced and semantically coherent chunks while avoiding costly global re-partitioning. For extraction, we concatenate $\mathrm{PreC}_i$ and $C_i$ with an extraction prompt $p_{\mathrm{ext}}$, and query the LLM: 
\begin{equation}\label{know-extr} 
\resizebox{.91\linewidth}{!}{$
K=\{k_{1},k_{2},\ldots,k_{n_K}\} =\bigcup_{i=1}^{N_C} {\rm LLM}\!\left(C_i,\mathrm{PreC}_i, p_{\mathrm{ext}}\right),
$}
\end{equation} 
where $n_K = |K|$ is the number of extracted items. Each extracted item is $k_i=\big(k_i^{\text{text}},\,k_i^{\text{ref}},\,k_i^{\text{entities}}\big)$, where $k_i^{\text{text}}$ is the relational statement, $k_i^{\text{ref}}$ is the supporting (unmodified) source span referenced by $k_i^{\text{text}}$, and $k_i^{\text{entities}}$ is the associated entity set (see Figure~\ref{fig:Knowledge} in Appendix~F for an example). Because adjacent windows share most prefix chunks, this strategy reduces boundary effects while remaining scalable; in practice, overlapping prefixes can also be cached to reduce repeated computation.

\paragraph{Knowledge Hypergraph Construction.} 
Each extracted knowledge item $k_i$ induces a hyperedge $(e_i^{H}, V_{e_i^{H}})$, where $V_{e_i^{H}}$ is the set of entities mentioned in $k_i$, and $e_i^{H}=\big(e_i^{H,\text{text}},\,e_i^{H,\text{ref}}\big)$ stores the hyperedge description and its supporting source span. Collecting all hyperedges yields 
\begin{equation}\label{hyperedge-entity-extr} 
E^{H}=\{e^{H}_{1}, e^{H}_{2}, \ldots, e^{H}_{n_K}\}, \qquad V=\bigcup_{i=1}^{n_K} V_{e_i^{H}}, 
\end{equation} 
where $V$ is the resulting entity set and $E^{H}$ is the hyperedge set. Each entity node $v=(v^{\text{name}}, v^{\text{text}})$ contains an entity name and an entity summary. To obtain robust entity summaries, we generate $v^{\text{text}}$ \emph{after} all hyperedges for the document have been constructed. 
If the number of incident hyperedges (hyperdegree) exceeds $\tau_s$, we prompt an LLM to synthesize a concise summary from the incident hyperedge texts; otherwise, we concatenate those texts directly.  Deferring summarization improves coverage and reduces early summarization bias compared with incremental summarization.  Finally, FlexStructRAG preserves explicit links from each hyperedge to its source span, enabling document grounding and evidence tracing during retrieval. The resulting hypergraph is $G^{H}=(V,E^{H})$. 

\paragraph{Knowledge Graph Construction.} 
While hyperedges capture higher-order interactions, binary relations remain frequent and often sufficient for many queries. We therefore construct a complementary knowledge graph on top of the hypergraph. We treat entities whose hyperdegree exceeds a threshold $\tau_e$ as \emph{anchors}. 
For each anchor, we collect its incident hyperedges and the neighboring entities appearing in those hyperedges, then prompt an LLM to extract explicit pairwise relations supported by the corresponding evidence spans. For each detected relation, we create an edge $e=(e^{\text{source}}, e^{\text{target}}, e^{\text{text}})$, where $e^{\text{source}}$ and $e^{\text{target}}$ are the incident entity nodes and $e^{\text{text}}$ is the relation description (optionally linked to the same supporting span as the triggering hyperedge). The resulting edge set $E=\{e_1,\ldots,e_{n_E}\}$ together with $V$ forms a standard knowledge graph $G=(V,E)$, where $n_E=|E|$ is the number of extracted edges. The thresholds $\tau_s$ (summarization) and $\tau_e$ (edge extraction) jointly control construction cost and graph density.

\paragraph{Meso-Level Structure-Aware Semantic Clustering (SSC).} 
LLM-based extraction yields relational primitives (edges and hyperedges), but long documents often scatter evidence across nearby spans, exhibit surface-form variation (synonymy/homonymy), and contain local extraction noise. Consequently, retrieving isolated primitives may lose vital context, whereas document-level aggregation risks introducing noise.

SSC mitigates this trade-off by clustering hyperedges (i.e., extracted n-ary relational statements) into intermediate, document-grounded semantic units. This meso-level granularity is intentionally between micro-level primitives (single hyperedges) and macro-level topic/document clusters, producing units that are context-rich yet still query-specific.

Each hyperedge $e_i^{H}$ is encoded as an embedding $e_{i}^{H,\text{emb}}$. Let $k_i^{\text{index}}$ denote the extraction order of the item that induces $e_i^{H}$ (Eq.~\ref{know-extr}). We define a pairwise distance that combines semantic dissimilarity (cosine distance) and positional separation in the extraction sequence: 

\begin{align}\label{sema-dis} 
&{\rm Dis}^{\text{pos}}_{ij}=\big|k^{\text{index}}_i-k^{\text{index}}_j\big|, \quad {\rm Dis}^{\text{cos}}_{ij}=1-\frac{\langle e_{i}^{H,\text{emb}},\, e_{j}^{H,\text{emb}}\rangle}{\|e_{i}^{H,\text{emb}}\|\,\|e_{j}^{H,\text{emb}}\|}, \notag\\ 
&{\rm Dis}_{ij}={\rm Dis}^{\text{cos}}_{ij}+\alpha\cdot{\rm Dis}^{\text{pos}}_{ij}, 
\end{align} 
where $\alpha$ controls the trade-off between semantic similarity and locality. 

We apply HDBSCAN~\cite{mcinnes2017hdbscan} to cluster hyperedges into $CH=\{ch_1, ch_2, \ldots\}$. HDBSCAN adapts to varying cluster densities and can label low-confidence hyperedges as noise under noisy extraction. Each cluster $ch_i=(ch_i^{\text{id}}, ch_i^{\text{text}})$ is assigned a unique identifier. The cluster reference text $ch_i^{\text{text}}$ is constructed by concatenating (in extraction order) the source spans referenced by hyperedges in the cluster, yielding document-grounded context that can be passed directly to the generator. 

Unlike approaches that cluster entity nodes or text chunks~\cite{edgerag2024,gutierrez2025ragmemorynonparametriccontinual,Wang2025ArchRAGAC}, SSC clusters relational statements (hyperedges). This directly aggregates higher-order relational evidence while retaining traceability to the supporting spans, enabling robust retrieval even when individual extracted items are incomplete. After SSC, each document $D_i$ yields three aligned structures: a hypergraph $G_i^{H}=(V_i,E_i^{H})$, a binary knowledge graph $G_i=(V_i,E_i)$, and a cluster set $CH_i$ composed of hyperedge-level semantic clusters. These structures jointly support flexible, multi-granular retrieval in the subsequent phase.

\subsection{Retrieval Strategy Phase} 
\begin{table*}[!tb]
	\centering
    \resizebox{.99\linewidth}{!}{$
   	\setlength{\tabcolsep}{0.88mm}{
		\begin{tabular}{lrrrrrrrrrrrr}
			\hline
			\multirow{2}{*}{Methods} & \multicolumn{3}{c}{Agriculture} & \multicolumn{3}{c}{CS} & \multicolumn{3}{c}{Legal} & \multicolumn{3}{c}{Mix} \\
			\cline{2-13}
			& EM & F1 & GE & EM & F1 & GE & EM & F1& GE & EM & F1 & GE \\
			\hline
			NaiveGeneration \cite{achiam2023gpt} & 9.38 & 15.38  & 39.80 & 15.63 & 24.56  & 45.78 & 12.50 & 23.88 & 42.82 & 10.16 & 17.48 & 36.22 \\
			
			StandardRAG \cite{lewis2020retrieval} & 18.75 & 28.75  & 51.88 & 28.13 & 38.14  & 56.40 & 27.73 & 38.51  & 54.98 & 35.16 & 47.40  & 61.85 \\
			
			LightRAG \cite{guo2024lightrag} & 17.58 & 25.51  & 46.67 & 17.77 & 27.18 & 47.35 & 19.92 & 29.13 & 45.20 & 31.84 & 42.84 & 56.86 \\
			
			GraphRAG \cite{edgerag2024} & 17.58 & 26.02 & 47.03 & 25.20 & 35.14 & 53.04 & 21.48 & 31.21 & 48.01 & 36.52 & 47.52  & 60.99 \\
			
			HippoRAG \cite{jimenez2024hipporag} & 13.28 & 19.42 & 43.44 & 19.34 & 28.51 & 48.13 & 18.75 & 28.75 & 45.94 & 28.71 & 38.02  & 52.83 \\
			
			RAPTOR \cite{sarthi2024raptor} & 16.60 & 24.60  & 47.68 & 24.22 & 34.15  & 53.23 & 21.29 & 32.38  & 49.79 & 37.50 & 49.17  & 62.75 \\
            Cog-RAG~\cite{hu2025cograg} &17.57    &24.72    &41.82      &25.39      &35.74      &48.44         &22.65      &32.97       &45.15      &29.88     &40.94     &48.92  \\
			  Hyper-RAG~\cite{feng2025hyper}   &17.57    &25.74    &42.88      &25.97      &36.00      &49.04         &23.04      &34.76       &47.28      &39.25     &49.91     &56.82  \\
			HyperGraphRAG \cite{luo2025hypergraphrag} & 29.10 & 39.23 & 57.83 & 35.94 & 45.30  & 60.57 & 31.84 & 42.29  & 57.19 & 44.34 & 55.84 & 66.89 \\
			\hline
			{\bf FlexStructRAG (ours) }    & \textbf{34.96} & \textbf{44.75}  & \textbf{60.59} & \textbf{41.02} & \textbf{51.15}  & \textbf{63.68} & \textbf{35.16} & \textbf{45.61} & \textbf{59.40} & \textbf{54.49} & \textbf{65.21}  & \textbf{73.09} \\
			\hline
			Improvement  & 5.86    & 5.52  &   2.76   &  5.08  & 5.85&  3.11  &  3.32  & 3.32   & 2.21   &  10.15  &   9.37  &   6.20  \\
			\hline
		\end{tabular}
        }
$	} 
\caption{UltraDomain results across four domains. Exact Match (EM), token-level F1, and GE (LLM-based generation evaluation) for nine baselines and FlexStructRAG. Bold indicates best per column; ``Improvement'' is relative to the best baseline (HyperGraphRAG) per domain.}
	\label{results}
\end{table*}

Given a query $q$, FlexStructRAG retrieves query-relevant evidence from the constructed knowledge base. Since knowledge is constructed per document to reduce spurious cross-document links, we unify the resulting structures at inference time to enable retrieval over the full corpus. Concretely, we merge per-document node, hyperedge, edge, and cluster sets into global collections: $\widetilde{V}= \bigcup^{N_D}_{i=1} V_i$, $\widetilde{E^{H}}=\bigcup^{N_D}_{i=1} E^{H}_i$, $\widetilde{E}= \bigcup^{N_D}_{i=1} E_i$, and $\widetilde{CH}=\bigcup^{N_D}_{i=1} CH_i$. On top of these global structures, FlexStructRAG provides four complementary retrieval modules. 

\paragraph{Entity Retrieval.}
We extract entities from the query using an entity-extraction prompt $p_{q_{\text{ext}}}$:
\begin{equation}\label{query-extr}
V_q = {\rm LLM}(p_{q_{\text{ext}}}, q).
\end{equation}
Each extracted entity is matched against $\widetilde{V}$ using a robust matching strategy: we first apply normalized string matching (lowercasing and punctuation removal). If no match is found, we consult an alias table constructed from LLM-generated entity summaries. We utilize exact string matching to retrieve relevant entity nodes.

\paragraph{Hyperedge Retrieval.}
Given the query embedding $q^{\text{emb}}$, we compute cosine similarity to each hyperedge embedding $e^{H,\text{emb}}$ for $e^{H}\in\widetilde{E^{H}}$.  We retain hyperedges with similarity above $\tau_{E_H}$ and return the top $N_{E_H}$ by similarity:
\begin{equation}\label{hyperege-retr}
\resizebox{.91\linewidth}{!}{$
f(E^H, q)=\operatorname{Top-}N_{E_H}\!\left(\left\{ e^{H}\in \widetilde{E^{H}} \mid {\rm sim}(q^{\text{emb}}, e^{H,\text{emb}})>\tau_{E_H} \right\}\right).
$}
\end{equation}
where ${\rm sim}(\cdot,\cdot)$ denotes cosine similarity. If fewer than $N_{E_H}$ hyperedges satisfy $\tau_{E_H}$, we return all of them.

\paragraph{Edge Retrieval.}
Given the query embedding $q^{\text{emb}}$, we compute cosine similarity to each edge embedding $e^{\text{emb}}$ for $e \in \widetilde{E}$. We retain edges with similarity above $\tau_E$ and return the top $N_{E}$ by similarity:
\begin{equation}\label{edge-retr}
\resizebox{.91\linewidth}{!}{$
f(E, q)=\operatorname{Top-}N_{E}\!\left(\left\{ e\in \widetilde{E} \mid {\rm sim}(q^{\text{emb}}, e^{\text{emb}})>\tau_{E} \right\}\right).
$}
\end{equation}
where ${\rm sim}(\cdot,\cdot)$ denotes cosine similarity. If fewer than $N_E$ edges satisfy $\tau_E$, we return all of them. For both edges and hyperedges, embeddings are computed from the relational text $e^{\text{text}}$ concatenated with the names of incident entities, so that semantic content and structural roles are jointly encoded.

\paragraph{Cluster Retrieval.} 
When queries require local discourse context beyond individual relations, we retrieve clusters containing the retrieved hyperedges.  Let $E^H_r \subseteq \widetilde{E^{H}}$ be the retrieved hyperedges and let $H(ch_i)$ denote the hyperedges assigned to cluster $ch_i$. We select clusters that contain at least one retrieved hyperedge: 
\begin{equation}\label{cluster-retr}
 CH_r=\big\{ch_i \in \widetilde{CH} \,\big|\, H(ch_i) \cap E^H_r \neq \emptyset \big\}. 
\end{equation} 
The cluster reference texts provide document-grounded context that can be passed directly to the generator. 

\paragraph{Retrieval Modes.} 
FlexStructRAG supports three configurable modes: \emph{hypergraph mode} (entity + hyperedge + cluster retrieval), \emph{graph mode} (entity + edge retrieval), and \emph{hybrid mode} (all modules). Unless otherwise stated, we use the hybrid mode as the default configuration.

\section{Experiments}\label{expe}
We evaluate FlexStructRAG with five research questions (RQ1--RQ5) covering effectiveness, ablations, parameter sensitivity, response quality, and efficiency. Additional analysis, case study, and prompt templates are provided in the appendices.

\subsection{Settings} 
\paragraph{Datasets.} 
We evaluate FlexStructRAG on four domains from the UltraDomain benchmark~\cite{qian2024memorag} following the protocol of LightRAG~\cite{guo2024lightrag}: Agriculture, Computer Science (CS), Legal, and a mixed-domain setting (Mix). Each domain contains 512 question--answer pairs curated in~\cite{luo2025hypergraphrag}. UltraDomain consists of long, knowledge-intensive documents that require multi-hop reasoning and precise evidence selection, making it suitable for evaluating structure-aware RAG systems. 

\paragraph{Baselines.} 
We compare against nine representative publicly available baselines:  \textbf{NaiveGeneration}~\cite{achiam2023gpt},
\textbf{StandardRAG}~\cite{lewis2020retrieval},
\textbf{RAPTOR}~\cite{sarthi2024raptor},
\textbf{LightRAG}~\cite{guo2024lightrag},
\textbf{GraphRAG}~\cite{edgerag2024},
\textbf{HippoRAG}~\cite{jimenez2024hipporag},
\textbf{Cog-RAG}~\cite{hu2025cograg},
\textbf{Hyper-RAG}~\cite{feng2025hyper},
and \textbf{HyperGraphRAG}~\cite{luo2025hypergraphrag}.
NaiveGeneration answers questions without retrieval, StandardRAG is the canonical chunk-based pipeline, and RAPTOR performs hierarchical chunk clustering. LightRAG, GraphRAG, and HippoRAG are graph-based retrieval methods, while Cog-RAG, Hyper-RAG, and HyperGraphRAG incorporate higher-order structure via hypergraphs. \paragraph{Evaluation Metrics.} We report Exact Match (EM), token-level F1, and Generation Evaluation (GE). EM measures exact string equivalence with the reference answer, F1 measures token overlap, and GE~\cite{hellobench} is an LLM-based metric that evaluates generation quality across seven dimensions. We report the average GE score following~\cite{luo2025hypergraphrag}. 

\paragraph{Implementation Details.} 
We use qwen-turbo-2025-07-15 as the backbone LLM and text-embedding-v3 for embedding computation. Unless otherwise specified, we set $ct_{\min}=500$, $ct_{\max}=600$, $g_{\max}=3$, $g_{\mathrm{overlap}}=2$, $\tau_s=9$, $\tau_e=3$, and $\alpha=0.1$. Retrieval parameters are $N_{E_H}=7$, $N_E=3$, $\tau_{E_H}=0.9$, and $\tau_E=0.9$. We choose hyperparameters based on sensitivity analysis (Section~\ref{para-sen} and Appendices). All experiments run on a Linux server with an NVIDIA GeForce RTX~4090 GPU.

\subsection{Experimental Results (RQ1)} 

Table~\ref{results} compares FlexStructRAG with nine baselines across four domains. FlexStructRAG achieves the best EM, F1, and GE in every setting. 

On Agriculture, FlexStructRAG attains 34.96 EM, 44.75 F1, and 60.59 GE, improving over the strongest baseline (HyperGraphRAG) by +5.86 EM, +5.52 F1, and +2.76 GE. On CS, FlexStructRAG reaches 41.02 EM, 51.15 F1, and 63.68 GE, exceeding HyperGraphRAG by +5.08 EM, +5.85 F1, and +3.11 GE. On Legal, which features more complex legal reasoning, FlexStructRAG remains consistently stronger than all baselines (35.16 EM, 45.61 F1, and 59.40 GE). The largest gains occur on Mix, where FlexStructRAG achieves 54.49 EM, 65.21 F1, and 73.09 GE, improving over HyperGraphRAG by +10.15 EM, +9.37 F1, and +6.20 GE. 

We attribute these gains to three design choices: (i) multi-granular retrieval over entities, binary relations, n-ary relations, and document-grounded clusters; (ii) structure-aware semantic clustering, which aggregates locally related relational evidence into query-usable context units; and (iii) truncated sliding-window extraction, which preserves bounded cross-chunk dependencies during knowledge construction.

\subsection{Ablation Study (RQ2)}
\begin{table}[!tb]
	\centering
	
	\begin{tabular}{lccc}
		\hline
		Methods & EM    & F1  & GE  \\
		\hline
		FlexStructRAG & \textbf{54.49} & \textbf{65.21} & \textbf{73.09} \\
		w/o SW & 52.15 & 63.25 & 71.95 \\
		w/o SSC & 46.09 & 57.60 & 67.37 \\
		w/o HR & 49.02 & 61.47 & 70.71 \\
		w/o EnR & 51.75 & 63.29 & 71.06 \\
		w/o ER  & 52.34 & 63.12 & 71.35 \\
		\hline
		w/o ER+SSC & 46.29 & 58.64 & 68.13 \\
		w/o ER+HR & 49.22 & 61.27 & 70.87 \\
		w/o ER+EnR & 51.95 & 62.93 & 71.19 \\
		\hline
		w/o HR+SSC  & 39.65 & 51.39 & 61.41 \\
		w/o HR+SSC+ER & 35.16 & 44.56 & 55.98 \\
		w/o HR+SSC+EnR & 35.74 & 47.22 & 58.44 \\
		
		\hline
	\end{tabular}
	\caption{Ablation on Mix domain. Impact of removing retrieval modules: Entity Retrieval (EnR), Edge Retrieval (ER), Hyperedge Retrieval (HR), and construction modules: Sliding Window (SW) and SSC on EM/F1/GE. ``w/o" denotes removal; combined removals test component complementarity.}
	\label{abla_study}
\end{table}    
\begin{figure}[!tb]
	\centering
	\subfigure[The impact of $N_{E_H}$]{
		\centering
		\includegraphics[scale=0.26]{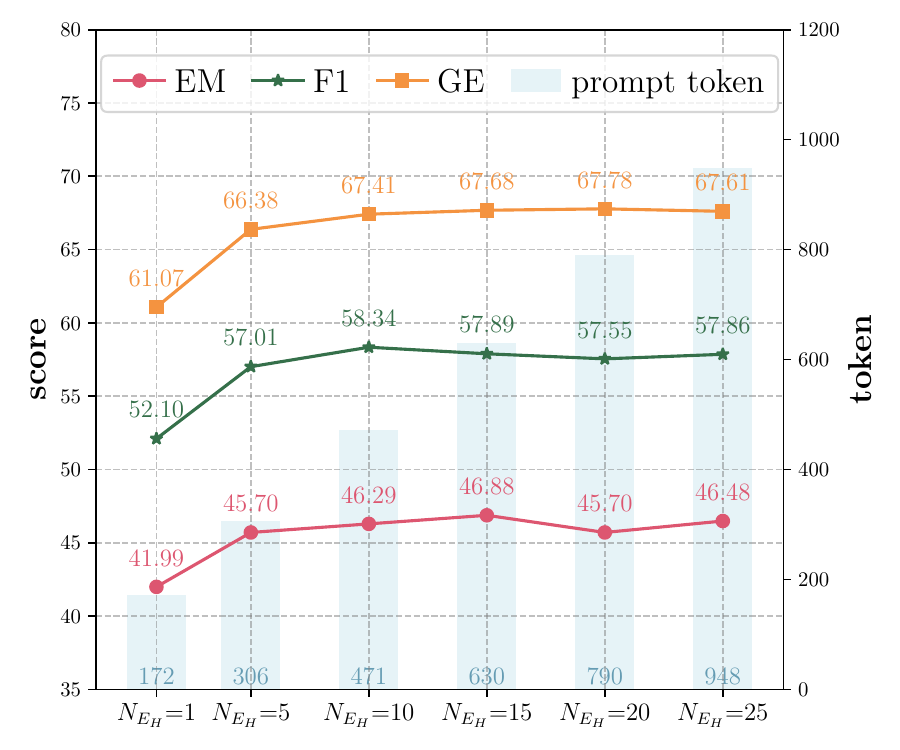}
		\label{fig:1}
	}
	\subfigure[The impact of  $N_E$]{
		\centering
		\includegraphics[scale=0.26]{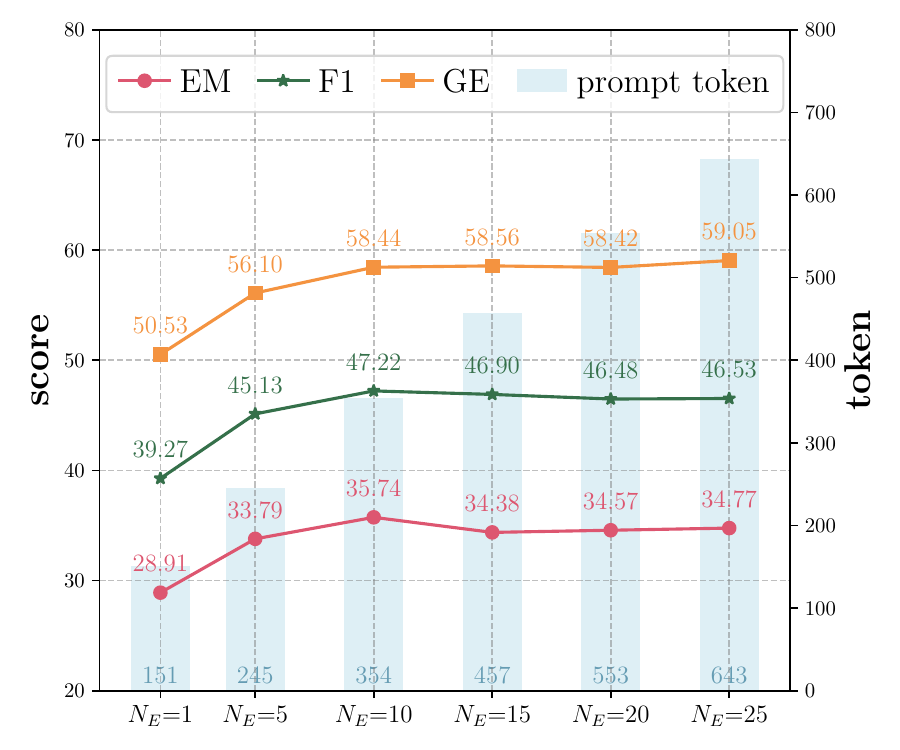}
		\label{fig:2}
	}
	
	\subfigure[The impact of   $\tau_s$]{
		\centering
		\includegraphics[scale=0.27]{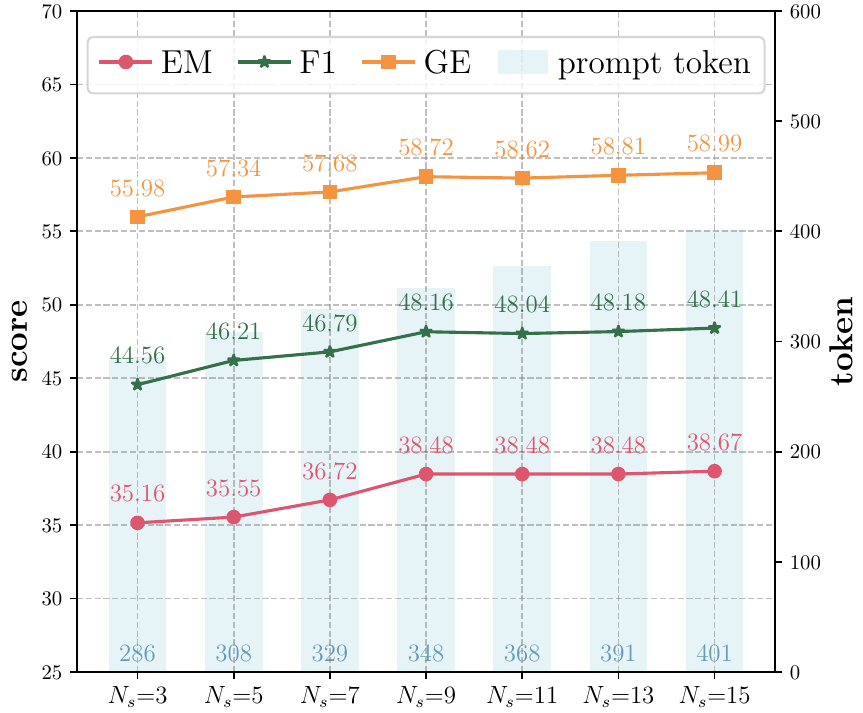}
		\label{fig:3}
	}
	\subfigure[The impact of  $\tau_e$]{
		\centering
		\includegraphics[scale=0.27]{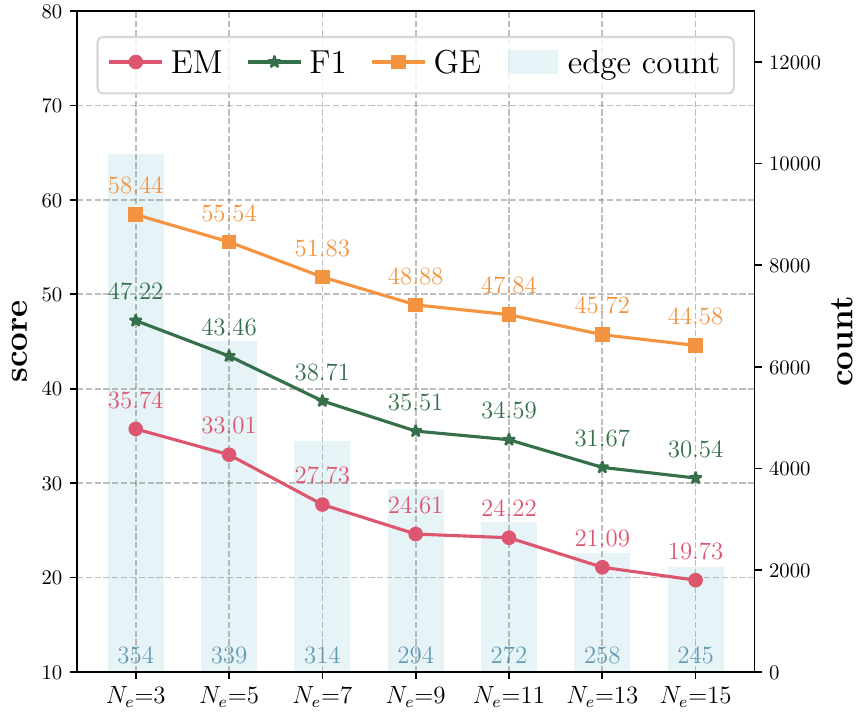}
		\label{fig:4}
	}	
	\caption{Parameter sensitivity analysis on the Mix domain. Performance variation with respect to (a) the maximum number of retrieved hyperedges $N_{E_H}$, (b) the maximum number of retrieved edges $N_E$, (c) the node summarization threshold $\tau_s$, and (d) the edge extraction threshold $\tau_e$.}
	\label{para-anal}
\end{figure}

\begin{table*}[!tb]
	\centering	
	\setlength{\tabcolsep}{4mm}{
		\begin{tabular}{lrrr|rrr}
			\hline
			\multicolumn{1}{l}{\multirow{2}{*}{Methods}} & \multicolumn{3}{c}{Construction}  & \multicolumn{3}{c}{Generation} \\ \cline{2-7} 
			\multicolumn{1}{c}{} & Prompt & Response & Time (s) & Prompt & Response & Time (s) \\ \hline
			LightRAG & 1,004,714 & 357,411 & 296 & 5,375 & 208 & 13.14     \\
			GraphRAG & 974,268  & 356,694 & 351 & 3,774 & 174 & 5.71     \\
			HippoRAG & 384,681 & 151,365 & 246 & 7,576 & 158 & 8.65     \\
			RAPTOR & 141,034 & 2,700 & 245 & 5,259 & 151 & 3.32     \\
            Cog-RAG &408,689   &558,829 &971   &13,205   &235   &12.95    \\
            Hyper-RAG&603,783   &989,861 &1,742   &19,235   &243   &6.51     \\
			HyperGraphRAG & 3,915,749 & 1,069,120 & 1232 & 12,221 & 179 & 4.44     \\ \hline
            {\bf FlexStructRAG (ours) } & 1,607,499 & 924,066 & 1062 & 1,097 & 211 & 3.18 \\
			\hline
		\end{tabular}
	}
	\caption{Comparison of computational cost (token consumption) and runtime (seconds) during the knowledge construction and answer generation phases. }
	\label{gene-cost}
\end{table*}

We conduct ablation experiments on the Mix domain to quantify the contribution of each component. We remove Entity Retrieval (EnR), Edge Retrieval (ER), Hyperedge Retrieval (HR), Structure-aware Semantic Clustering (SSC), and the Sliding-Window mechanism (SW), and also evaluate combinations. Results are reported in Table~\ref{abla_study}. 

\paragraph{Effect of Entity Retrieval (EnR).} 
Removing EnR degrades performance (EM$\downarrow$2.74, F1$\downarrow$1.92, GE$\downarrow$2.03), indicating that explicit entity grounding improves evidence alignment and generation quality. The degradation is larger when EnR is removed together with ER or HR. 

\paragraph{Effect of Hyperedge Retrieval (HR).} 
Removing HR reduces performance substantially (EM$\downarrow$5.47, F1$\downarrow$3.74, GE$\downarrow$2.38), highlighting the importance of retrieving n-ary relational evidence. Combined removals (e.g., HR+SSC) lead to further degradation, suggesting that higher-order relations and meso-level context are complementary. 

\paragraph{Effect of Structure-Aware Semantic Clustering (SSC).} 
SSC is the most influential component. Removing SSC yields the largest drop among single-component ablations (EM$\downarrow$8.40, F1$\downarrow$7.61, GE$\downarrow$5.72). SSC provides meso-level aggregation that compensates for earlier chunking-induced fragmentation.

\paragraph{Effect of Edge Retrieval (ER).} 
Removing ER decreases performance (EM$\downarrow$2.15, F1$\downarrow$2.09, GE$\downarrow$1.74), indicating that binary relations remain valuable even when higher-order evidence is available. The drop increases when ER is removed jointly with HR or SSC. 

\paragraph{Effect of Sliding Window (SW).} 
Removing SW also consistently hurts performance (EM$\downarrow$2.34, F1$\downarrow$1.96, GE$\downarrow$1.14), validating that incorporating bounded contextual dependencies improves extraction quality and downstream retrieval. 

Overall, all components contribute to the final performance, with SSC providing the largest marginal benefit.

\subsection{Parameter Sensitivity Analysis (RQ3)}\label{para-sen}

We study the trade-off between effectiveness and prompt cost by varying $N_{E_H}$, $N_E$, $\tau_s$, and $\tau_e$ on the Mix domain (Figure~\ref{para-anal}). Overall, increasing retrieval breadth improves performance only up to a moderate range, after which gains saturate while prompt cost continues to grow. 

Specifically, increasing the maximum number of retrieved hyperedges $N_{E_H}$ or edges $N_E$ yields initial accuracy improvements but exhibits diminishing returns beyond moderate values, while prompt cost grows approximately linearly.  For entity-only retrieval, increasing the entity summary threshold $\tau_s$ slightly improves performance at the expense of longer prompts, whereas overly small $\tau_s$ leads to aggressive summarization and information loss. Finally, a larger edge extraction threshold $\tau_e$ reduces the number of extracted edges; in edge-only settings this sharply degrades performance due to insufficient relational coverage.

\subsection{Analysis of the Response Quality (RQ4)}

\begin{figure}[!tb]
	\centering
	\includegraphics[scale=0.42]{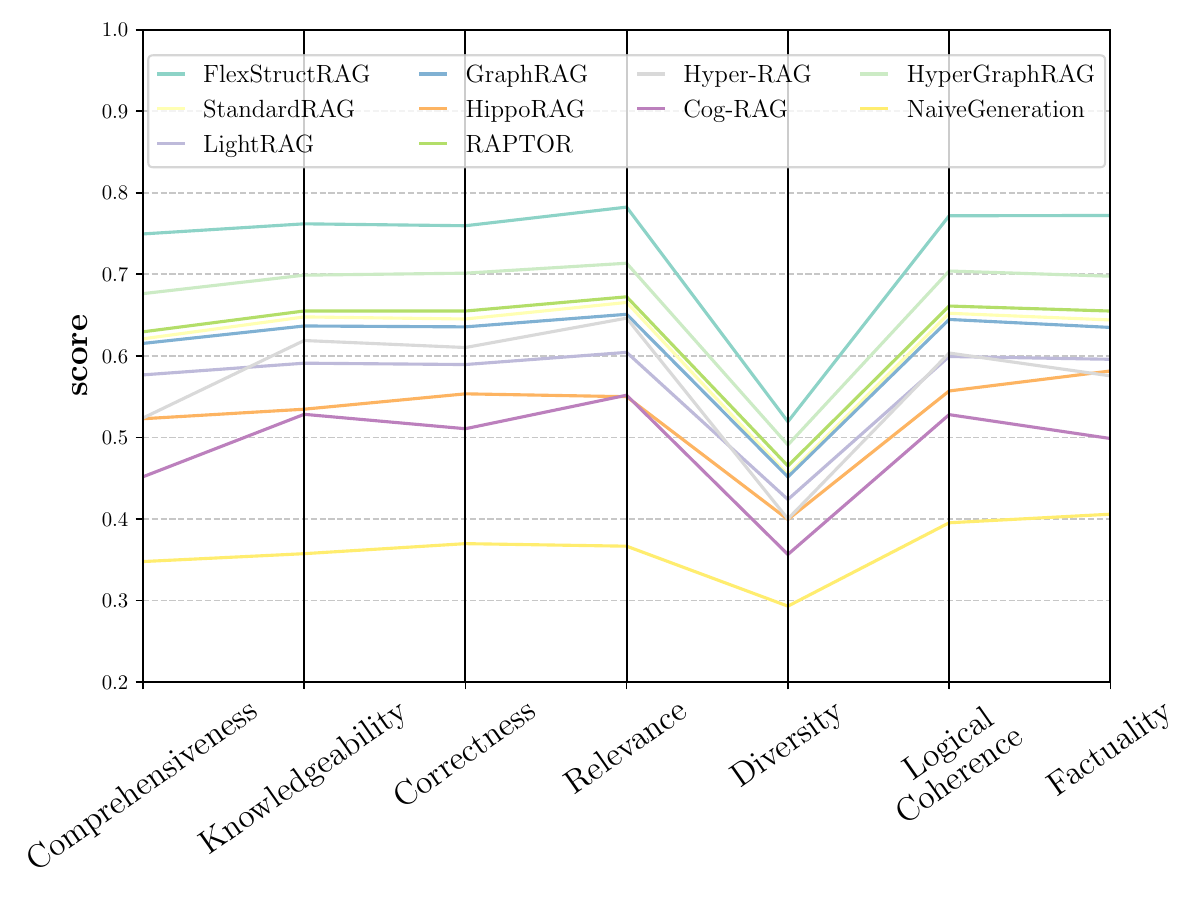}	
	\caption{Generation-quality breakdown (GE) across seven evaluation dimensions: Comprehensiveness, Knowledgeability, Correctness, Relevance, Diversity, Logical Coherence, and Factuality. Results compare FlexStructRAG with baseline methods on the Mix domain.}
	\label{gene_qual} 
\end{figure}

We further analyze response quality using GE across seven dimensions: Comprehensiveness, Knowledgeability, Correctness, Relevance, Diversity, Logical Coherence, and Factuality. Figure~\ref{gene_qual} shows that FlexStructRAG achieves the highest overall GE score among all methods and performs strongly across all dimensions. In particular, it reaches approximately 0.8 on Relevance and Logical Coherence, indicating effective evidence alignment and consistent reasoning. StandardRAG and LightRAG exhibit stable but uniformly lower scores. GraphRAG and HippoRAG perform competitively on Correctness and Relevance but trail on Diversity. RAPTOR shows a pronounced reduction in Diversity (around 0.4), which is consistent with broader hierarchical aggregation introducing less varied evidence. HyperGraphRAG scores highly on Logical Coherence and Factuality, but remains below FlexStructRAG overall.

\subsection{Analysis of Efficiency and Computational Cost (RQ5)}  

We compare FlexStructRAG with seven baselines (LightRAG, GraphRAG, HippoRAG, RAPTOR, Cog-RAG, Hyper-RAG, and HyperGraphRAG) in terms of token usage and runtime for both knowledge construction and answer generation. Table~\ref{gene-cost} reports prompt tokens, response tokens, and wall-clock time. FlexStructRAG incurs higher one-time construction cost (1,607,499 prompt tokens and 924,066 response tokens), reflecting the one-time expense of extracting relations, constructing multiple structures, and forming clusters. However, this cost is substantially lower than HyperGraphRAG (3,915,749 prompt tokens and 1,069,120 response tokens). At inference time, FlexStructRAG is notably more efficient: it requires only 1,097 prompt tokens per query, compared with 12,221 for HyperGraphRAG, 19,235 for Hyper-RAG, 13,205 for Cog-RAG, and 5,259 for RAPTOR. This reduction suggests that FlexStructRAG shifts computation to offline construction and produces more compact, query-aligned evidence for generation, thereby lowering online overhead.
\section{Conclusion}\label{conc}
We present FlexStructRAG, a structure-aware RAG framework that unifies graph, hypergraph, and document-grounded retrieval. By jointly modeling binary relations, n-ary relations, and meso-level semantic clusters, it captures complex relational evidence while preserving document grounding. 

Compared with existing approaches, FlexStructRAG introduces four key advances: (1) a unified multi-unit knowledge base that enables inference-time mode switching without reindexing; (2) hyperedge-centered structure-aware semantic clustering with positional locality to bridge symbolic relations and discourse context; (3) dynamic partitioning with truncated sliding-window extraction, together with span-level provenance links, to reduce chunking distortion while keeping construction bounded, linear-time complexity and cacheability; and (4) feature toggles that subsume graph-only, hypergraph-only, and structure-only variants, enabling controlled comparisons. Overall, FlexStructRAG provides a scalable and effective solution for complex knowledge-intensive generation tasks in real-world settings.

\newpage
\bibliographystyle{named}
\bibliography{reference}

\end{document}